\documentclass[prb,aps,onecolumn,groupedaddress,floats,showpacs,final,12pt]{revtex4}
\usepackage{graphicx}
\usepackage{dcolumn}
\usepackage{bm}
\usepackage{color}
\usepackage{ulem}
\definecolor{blue}{rgb}{0.3,0.3,0.9}

\def\beq{\begin{eqnarray}}
\def\eeq{\end{eqnarray}}
\def\i{{\rm i}}

\begin{document}

\author{Victor Fleurov$^1$ and Anatoly Kuklov$^2$}
\affiliation{$^1$ Raymond and Beverly Sackler Faculty of Exact Sciences, School of Physics and Astronomy, Tel-Aviv University -
Tel-Aviv 69978, Israel }
\affiliation{$^2$ Department of Physics and Astronomy, CSI, and the Graduate Center of CUNY, New York.}

\title{Phases and phase transitions of Bose condensed light: phase separation efect.}

\date{\today}
\begin{abstract}
Recent realization of Bose-Einstein condensation of light in 2D provides a new platform for studying novel phases and phase transitions. The combination of low effective mass of the confined light and the presence of the dye molecules with randomly oriented directions of the dipolar transition engages a competition between disorder and the tendency to forming algebraic off-diagonal order. The phase diagram of possible phases is constructed at the mean field level.  One of the phases  is the condensate of photon pairs induced solely by the orientational disorder. Such a {\it geometrical } mechanism of pairing has no analogy in other systems. It is also found that the photon condensantion can proceed as a phase separation effect leading to a non-uniform distribution of the condensate and dye molecules.
\end{abstract}


\maketitle

\section{Introduction}
About 25 years ago a new revolution in physics took place -- three groups [\onlinecite{BEC1,BEC2,BEC3}] realized Bose Einstein condensates of ultracold atoms.
This has initiated a search for new phases of matter culminating in demonstrating Bose - Mott insulator quantum phase transition [\onlinecite{Mott}]. Since then many more proposals of exotic novel strongly interacting phases of matter have been put forward and realized experimentally (see in Ref.[\onlinecite{RMP}]). Among the exciting possibilities is the bosonic fermionization [\onlinecite{Tigran}] induced by artificial spin-orbit coupling [\onlinecite{Spielman}].  Realization of interacting topological insulators with ultracold atoms also appears to be within the reach [\onlinecite{Bloch}]. To great extent these achievements have stimulated the development of strongly interacting photonics where the role of atoms is played by photons (see in Ref.[\onlinecite{Iacopo}]).

Creation of Bose-Einstein condensate (BEC) of light, Refs.[\onlinecite{Nature,Schmitt,NS14,MN15,GPO18}], has opened up a new chapter in the search for strongly interacting phases of light. In these experiments the thermalization of light is achieved through absorption and reemission of photons by dye molecules which represent a thermal bath -- thanks to their  manifolds of rovibrational states exchanging energy with the solvent [\onlinecite{KK15},\onlinecite{Axel}].
The equilibration of light occurs on much faster scale than the photon escape. Thus, for all practical purposes the emerging phase of light becomes a thermodynamical phase where photons can be characterized by finite chemical potential -- due to a significant difference between temperature and photon energies.

The most recent exciting development in this field includes realization of the lattice consisting of micro photonic condensates with the Josephson interaction between them [\onlinecite{Weitz2017}]. It clearly opens up a pathway toward realization of strongly interacting phases of light -- by means of independent tuning of the Josephson tunneling amplitude and onsite interaction.

At this juncture it is important to mention that there are some similarities and significant differences between the dye molecules in the resonator [\onlinecite{Nature,Schmitt,MN15,GPO18}]  and excitons in the condensed matter systems [\onlinecite{Blatt,Keldysh}] (see also in Ref.[\onlinecite{Moskalenko}]).  In the exciton-polaritonic materials the effective interaction between photons is induced through a direct exciton-exciton scattering -- thanks to the linear and coherent coupling between exciton and photon branches of the spectrum. Accordingly, the Gross-Pitaevskii (GP) equation including two polarizations of polaritons is straightforward to derive, Ref.[\onlinecite{Keeling-08}]. In the systems [\onlinecite{Nature,Schmitt,MN15,GPO18}] the dye ensemble is disordered and, thus, it cannot be coherently coupled to photons. [The coherency and interaction between dye molecules via exchanging near field 2D-photons was discussed in[\onlinecite{Sela-14}]].
An option to include photon-photon interaction has been suggested in Ref.[\onlinecite{NS14}] -- by taking into account the Kerr effect. Another option is the thermal lensing effect in the dye subsystem [\onlinecite{Nature,Weitz2017}].

The experimentally achieved condensation of light is characterized by weak effective interaction between photons [\onlinecite{Nature,Schmitt,MN15,GPO18}]. However, even infinitesimally small interaction changes dramatically properties of BEC (see in Ref. [\onlinecite{Landau_9}]). According to Ref. [\onlinecite{Nature}], the dimensionless interaction constant $\tilde{g}\sim 10^{-3}$ and typical photon numbers $N \sim 10^5$, with
the condensate size $D \sim 10-20\mu$m. This corresponds to the healing length $l_h \approx 1/\sqrt{\tilde{g}n_{\rm ph}}\approx 1-2\mu$m, where $ n_{\rm ph}\sim N/D^2$ is the 2D density of the photonic condensate. This estimate shows that, despite the smallness of the interaction, the healing length is much smaller than the condensate size and, thus, the GP equation is a relevant description of the system.

As will be discussed below, to achieve algebraic condensation of photons in 2D at finite temperature $T$ it is important to have anisotropy of the photon-photon interaction. In this respect, the thermo-optical effect which is insensitive to polarization [\onlinecite{Nature,Weitz2017}] cannot provide such an anisotropy.
Here we will address this aspect by including the anisotropic photon-photon interaction induced by the dye molecules each represented as a two level system (TLS) with randomly oriented vector $\vec{d}$ of the dipolar transition  between ground and excited states.  [In this analysis the role the dye sub ensemble in the thermalization of photons is set aside].  Furthermore, the limit of low density of the dye molecules is considered so that their interaction through near field photons, as discussed in Ref.[\onlinecite{Sela-14}] , can be safely ignored. The analysis is conducted for uniform condensate in the thermodynamical limit.

The focus of the study is on the {\it orientational} disorder of the field $\vec{d}$ in the dye ensemble.
We argue that it should result in a new phase of the photonic matter -- the {\it geometrically} paired photonic superfluid (PSF). More specifically, such a pairing is induced by spatial randomness in $\vec{d}$  so that no algebraic order can be detected in the one-photon density matrix, while the two-photon density matrix demonstrates the algebraic off-diagonal long range order (ODLRO). This result was first introduced in  Ref.[\onlinecite{FK18}] while here we devote a special attention to the derivation of the GP equation with an explicit Berry term. It is also pointed out here that allowing the TLS centers to change their positions due to the interaction with the PSF can lead to the phase separation effect, implying that the condensation can proceed simultaneously with the formation of non uniformity. Thus the full phase diagram includes the phase separation effect too.

Our paper is organized as follows. In Sec.\ref{order_disorder} the relevant variables are introduced. The Hamiltonian and the effective action are discussed in Sec.\ref{sec:H}. Symmetries of the phases are analyzed in Sec.\ref{sec:sym}, with the geometrically paired PSF introduced in Sec.\ref{main}. The phase separation effect is addressed in Sec. \ref{sec:PS}. Finally, the overview and discussion is presented in the Sec.\ref{Disc}.

\section{Order and disorder parameters}\label{order_disorder}

The order parameter of light is a complex vector $\vec{E}=\vec{E}(\vec{x},z)$ representing the amplitude of electric field in the rotating wave approximation (RWA), with $\vec{x}=(x,y)$ and $z$ being coordinates along and perpendicular to the XY plane of the resonator [\onlinecite{Nature}], respectively. The geometry of the experiment [\onlinecite{Nature}] selects a single longitudinal mode $\phi=\phi(z)$ along the resonator Z-axis so that the field can be represented as $\vec{E}=\phi(z) \vec{\psi}(\vec{x})$, where the complex vector $\vec{\psi}=(\psi_x, \psi_y)$ accounts for the transverse order and its long wave fluctuations in the XY-plane of the resonator. [There is also the $E_z$ component present insuring the divergenless nature of the photonic field
as $\nabla_zE_z+ \vec{\nabla} \vec{\psi} \phi(z)=0$].
We will be using the normalization in which $\vec{\psi}^\dagger \vec{\psi}$ is the operator of the 2D density of photons. Then, in the standard SI units
\beq
\phi(z)= \sqrt{\frac{2\hbar \omega_0}{\varepsilon_0L_z}}\sin(q_0 z),
\label{phi}
\eeq
[\onlinecite{com}], where $q_0=\omega_0/c$ is the wavevector of the standing wave, with $c$ being speed of light  and $L_z= 7\pi/q_0$ is the "height" of the resonator [\onlinecite{Nature}],  and $\varepsilon_0$ denotes the background electric permitivity of the media. The goal is to obtain the effective action for the amplitude $\vec{\psi}$ in the long wave limit by eliminating TLSs and integrating out the direction along $Z-$axis.

The TLS molecules can be accounted for by a coarse grained field of their dipolar transitions $\vec{d}=\vec{d}(\vec{x},z)=(d_x,d_y,d_z)$ responsible for absorbing and emitting photons. Such a variable emerges because the dye molecules are characterized by low symmetry so that the dipolar transition is non-degenerate, (see in Ref.[\onlinecite{Schmitt}])  and, thus, is described by a specific direction determined by spatial orientation of the molecule.

In order to derive the effective low energy description for 2D vector field $\vec{\psi}(\vec{x})$ it is necessary to project the 3D field $\vec{d}$ on the XY plane. Such a projection involves the integration of various tensors formed by $\vec{d}$ over $z$. Once projected along the XY plane of the resonator, this field plays the role of essentially a static and random gauge field.  It is also important to note that the orientational disorder of $\vec{d}$ precludes formation of the condensate of polariton [\onlinecite{EP}] (see in Ref.[\onlinecite{EP_RMP}]). Despite that, there is a novel feature emerging -- the geometrical pairing of photons due to the orientational disorder. This aspect of the system  is one of the main focuses of this work.

\section{The Hamiltonian and the effective action}\label{sec:H}

The Hamiltonian $H$ consists of three parts
\beq
H=H_{\rm ph} + H_{\rm TLS} + H_{\rm int},
\label{Hfull}
\eeq
where the first term accounts for the free photons inside the cavity; the second term describes the TLSs and the last one stands for the interaction between the photons and TLSs. Explicitly,
\beq
H_{\rm ph}=\int d^2x \left[ \frac{\hbar^2}{2m} \nabla_i \psi^\dagger_j \nabla_i \psi_j - \mu_0 \psi^\dagger_j\psi_j\right],
\label{U0}
\eeq
where $m=\hbar \omega_0/c^2$ stands for the effective mass of photons (induced by the dimensional quantization), and $\mu_0$ is the effective chemical potential of photons. The summation over the repeated coordinate indices (X,Y) here and below is implied.

The interaction can be written in terms of the photonic electric field $\vec{E}$ and the TLSs in the minimal form within the Rotating Wave Approximation (RWA) as
\beq
H_{int}= - \sum_{\vec{x},z} \left[ \vec{d} \vec{E} \sigma^+ +H.c.\right] ,
\label{int}
\eeq
where the summation runs over the locations of the TLSs and $\sigma^{+},\ \sigma^-$ are the Pauli matrices describing absorption and emission of photons, respectively, by each TLS located at the spatial points $(x,y,z)$ .[The component $E_z \sim \vec{\nabla}\vec{\psi}/q_0$ is ignored in Eq.(\ref{int}) in the limit of small momenta along X,Y directions]. The TLS energy can be accounted for by
\beq
H_{TLS} =\sum_{\vec{x},z} \delta \cdot \sigma_z,
\label{zigma}
\eeq
with $\delta = \epsilon_0 - \hbar \omega_0>0$ standing for the detuning of the TLS energy $\epsilon_0$ from the energy $\hbar \omega_0$ of the condensed photonic mode, and $\sigma_z$ being the Pauli matrix.

 Since we excluded vibrons from the consideration, the quantity $\epsilon_0$ should be attributed to Zero Phonon Line of the electronic transition in the dye molecule. In general, however, vibrons can change this interpretation. In what follows $\delta$ will be considered as a free parameter.

It is important to note that the Hamiltonian (\ref{Hfull}) conserves the total amount of photons and TLS molecules in their excited state. This can be explicitly seen after calculating the time derivative $\dot{\rho}=[\rho,H]/(i\hbar)$ of the excitation density operator
\beq
\rho (\vec{x},t)=\vec{\psi}^\dagger(\vec{x})\vec{\psi}(\vec{x}) + \sum_{\vec{x}_i,z_i,} \frac{1}{2} \sigma_z(\vec{x}_i,z_i) \delta^{(2)}(\vec{x}-\vec{x}_i)
\label{conserv}
\eeq
which leads to the 2D continuity equation
\begin{equation}
\frac{\partial \rho}{\partial t} + \vec{\nabla} \vec{J} =0,
\label{continuity}
\end{equation}
where
\begin{equation}
\vec{J} = \frac{\hbar}{2mi}\left[ \psi^\dagger_j \vec{\nabla} \psi_j - c.c.\right]
\label{J}
\end{equation}
is the photon current 2D density; $\delta^{(2)}(\vec{x}-\vec{x}_i)$ is 2D delta-function defined in the space $\vec{x}=(x,y)$.

It is important to emphasize that, in the absence of a direct exchange interaction between TLS molecules, the energy transfer is carried by photons only while storage of the energy is due to both -- photons and excitations of TLSs. The impact of the conservation (\ref{conserv}) on the mean field dynamics will be considered in detail in the Appendix A.

\subsection{Free energy}

In the presence of macroscopic occupation of photons forming the field $\vec{\psi}$, the quantum nature of $\vec{\psi},\, \vec{\psi}^\dagger$ can be safely ignored. Such an approach is the basis for describing superfluids by classical fields  $\vec{\psi},\, \vec{\psi}^*$ within the Gross-Pitaevskii equation (see in Refs.[\onlinecite{Landau,BK}]). This method is applicable in 2D at finite $T$ as well (see in Ref.[\onlinecite{BK}]) despite the absence of the true ODLRO
which is replaced by algebraic off diagonal order (and which we will be loosely referring to as "condensate").

In the presence of the condensate, the TLS contribution to the partition function can be calculated explicitly. Indeed, the thermal operator $\exp(-\beta (H_{TLS}+H_{int}))$, with $\beta=1/T$, can be represented as a product $\prod_i \exp(-\beta H_i)$ over each TLS where $H_i$ is the contribution from the $i$-th TLS to the terms (\ref{zigma},\ref{int}).
Then, using the identity $\exp(\vec{B}\vec{\sigma})= \cosh(|\vec{B}|) + \sinh(|\vec{B}) \vec{B}\vec{\sigma}/|\vec{B}|$ for Pauli matrices $\vec{\sigma}$ and real vector $\vec{B}$ and keeping in mind that that all the components of $\vec{\sigma}$ are traceless, the contribution to the free energy $U(\vec{\psi}, \vec{d}\,)= -T\ln {\rm Tr} [\exp(-\beta (H_{TLS} + H_{int}))]$, becomes
\beq
U_\psi([\vec{\psi}])= -T\sum_{x,y,z} \ln [2\cosh\left(\beta \epsilon\right)],
\label{Uwang}
\eeq
where
\beq
\epsilon=\sqrt{\delta^2 + |\vec{d}\vec{E}|^2}
\label{eps}
\eeq
is the half of the energy difference between the excited and the ground states of a TLS subjected to the classical field $\vec{E}$. It is worth mentioning that Eq.(\ref{Uwang}) is essentially the same as the one obtained in Ref.[\onlinecite{Wang}], although derived from a different perspective.

While considering low energy properties, it is reasonable to resort to a coarse grained description and, thus, to replace the summation $\sum_{x,y,z} ... $ over locations of TLSs by integration $\sum_{x,y,z} ... \to \int dz dy dx n ...$ where $n=n(x,y,z)$ stands for the coarse grained density of TLSs (which is not necessarily uniform).

At this point a comment about a possible degeneracy of the TLS transition is in order. If the TLS molecules were fully symmetric, the dipolar transition would be triple degenerate. Accordingly, each component of $\vec{E}$ would see no preferential orientation of the molecules, and a contribution to free energy of one molecule exposed to the condensate of light would depend on the product $d |\vec{E}| $, where $d = |\vec{d}|$.  Then, the free energy $U$, Eq.(\ref{Uwang}), would depend only on the modulus $|\vec{E}|$. In other words, the term $|\vec{d}\vec{E}|^2$ in Eq.(\ref{Uwang}) should be replaced by $d^2|\vec{E}|^2$. Accordingly, the effective action (represented as a Landau expansion [\onlinecite{Landau}]) would depend on $|\vec{\psi}|^2 $ and its higher powers. This implies the O(4) symmetry of the effective action. It is important that in 2D symmetries higher than O(2) preclude condensation at any finite temperature even in the algebraic sense [\onlinecite{Shenker}]. This aspect has been emphasized in the context of polariton condensation in Ref.[\onlinecite{Rubo-07}] and remains valid in the case under consideration too. Thus, in our analysis it is important that the TLS transition is not triple degenerate.

Here we consider the limit $|\delta| >>  |\vec{d}\vec{E}|$ validating the separation of fast and slow variables. Thus, $U$ in Eq.(\ref{Uwang}) can be expanded in $1/\delta$. The quadratic and quartic in $\vec{\psi}$ terms become
\beq
U_2=  \int d^2x [ -d_{ij}  \psi^*_i\psi_j] ,
\label{Fd}
\eeq
and
\beq
 U_4=  \int d^2x d_{ijkl} \psi^*_i \psi^*_j\psi_k\psi_l ,
 \label{Fd4}
\eeq
respectively. Here the tensors $d_{ij}=d_{ij}(x,y)$ and $d_{ijkl}=d_{ijkl}(x,y)$ are
\beq
d_{ij} = c_2 \int dz  n d_i d_j|\phi (z)|^2,
\label{dij}
\eeq
\beq
d_{ijkl}= c_4 \int dz n d_i d_j d_k d_l|\phi (z)|^4,
\label{dijkl}
\eeq
where $\phi (z)$ is defined in Eq.(\ref{phi}) and
\beq
c_2&=&  \frac{\tanh(\beta\delta)}{2\delta} ,
\label{c_2} \\
c_4 &=& \frac{\tanh(\beta\delta)}{8\delta^3} - \frac{1}{8T\delta^2\cosh^2(\beta\delta)}.
\label{c_4}
\eeq
It is worth mentioning that $c_4  >0$ and varies from $c_4=1/(8 \delta^3)$ for $\delta >>T$ to $c_4=1/(12T^3)$ in the opposite limit.

The summations in Eqs.(\ref{Fd},\ref{Fd4}) run over the X,Y directions only (because $\vec{\psi}$ has only X,Y components). Thus, $d_{ij}$ and $d_{ijkl}$ become 2D tensors. Then, it is convenient to introduce the representation
\beq
d_{ij} = \mu_{||} n_i n_j + \mu_{\perp}m_i m_j
\label{nmrep}
\eeq
in terms of the local frame where $d_{ij}$ is diagonal and $\mu_{||}, \mu_\perp$ are its eigenvalues. The unit vectors $\vec{n},\vec{m}$ can be conveniently represented as $\vec{n}= (\cos\theta, \sin\theta)$, $\vec{m} = (\sin\theta, -\cos\theta)$ by the angle $\theta=\theta(x,y)$ with respect to the X-axis. These vectors satisfy the orthogonality condition $\vec{n} \cdot \vec{m}=0$ so that the field $\vec{\psi}$ can be expanded as
\beq
\vec{\psi} = \vec{n} \Phi + {\rm i} \vec{m}\Phi_\perp,
\label{Phi}
\eeq
where $\Phi, \Phi_\perp$ are complex coordinates of $\vec{\psi}$ in the local frame $(\vec{n},\ \vec{m})$. Accordingly, the free energy (\ref{Uwang}) $U_\psi = U_2 + U_4$ together with the free photonic part (\ref{U0}) becomes
\begin{widetext}
\beq
U=\int d^2x \Big[\frac{\hbar^2}{2m}(|\vec{\nabla}\Phi_\perp + i \nabla \theta \Phi |^2 &+& |\vec{\nabla}\Phi + i \vec{\nabla}\theta \Phi_\perp|^2 )  
- (\mu_0 + \mu_{||}) |\Phi |^2 - (\mu_0 + \mu_{\perp}) |\Phi_\perp |^2\Big] \nonumber \\
&&+ U_4,
\label{Upsi}
\eeq
\end{widetext}
where the quartic term $U_4$ has been defined in Eqs.(\ref{Fd4}) and (\ref{dijkl}). Its specific form in terms of  the representation (\ref{Phi}) will be discussed later. The energy (\ref{Upsi}) represents the potential part of the full GP action $S$ derived in Appendix \ref{AppA}, Eq.(\ref{Stot}).

\section{Symmetries of the photonic condensate}\label{sec:sym}

In most cases the TLS transition is non-degenerate [\onlinecite{Moodie-17}]. The case of a double degenerate transition is qualitatively the same as the one considered below as long as the orientation of $\vec{d}$ is not aligned with the XY plane. 
In this section, we will first consider the case of full isotropy of the tensors $d_{ij}, d_{ijkl}$.

\subsection{Emerging isotropy}

It is natural to assume that the integration along Z-direction in Eqs.(\ref{dij}) and (\ref{dijkl}) returns isotropic tensors -- because the distance $ L_z$ (which is $\approx 2\mu$m) is much larger than the inter TLS separation (of the order of few nm). This allows for averaging over isotropic 3D orientations of $\vec{d}(x,y,z)$ in Eqs.(\ref{dij}) and (\ref{dijkl}), which gives
\beq
d_{ij} =\mu_\| \delta_{ij},
\label{iso}
\eeq
where $\mu_\| = \mu_\perp = \frac{c_2\vec{d}^2}{3} \int dz n |\phi|^2$. The isotropic quartic tensor $d_{ijkl}$ becomes
\beq
 d_{ijkl}=  g \left( \delta_{ij}\delta _{kl} + \delta_{ik}\delta_{jl}  + \delta_{il}\delta_{jk}\right),
\label{iso2}
\eeq
where $g=\frac{c_4\vec{d}^4}{15}\int dz n |\phi|^4$. It is worth mentioning that in the Kronecker symbols the indices run over $X,Y,Z$ directions while the summation in Eq.(\ref{Fd4}) is performed only over $X,Y$.  Accordingly, the substitution of Eq.(\ref{iso2}) into Eq.(\ref{Fd4}) gives
\beq
U_4 = \int d^2x\,  g\left(2|\vec{\psi}|^4 + (\vec{\psi}^*)^2(\vec{\psi})^2\right), 
\label{U44}
\eeq
or in terms of the representation (\ref{Phi}):
\beq
U_4=\int d^2x g\Big[3(|\Phi|^4 + |\Phi_\perp|^4) +4|\Phi|^2|\Phi_\perp|^2 -(\Phi^{*2}\Phi^{2}_\perp +c.c. )\Big] . %
\label{U4P}
\eeq
At this point it is important to note that, given $\mu_{||}=\mu_\perp$ and without the last term in $U_4$, Eq.(\ref{U44}), the symmetry of the free energy (\ref{Upsi}) is O(4). In this case, the vectors $\vec{n}$ and $\vec{m}$ in the representation (\ref{Phi}) can be gauged away and taken as aligned with the XY axes. Thus, the gauge field $\vec{\nabla}\theta=0$ in Eq.(\ref{Upsi}). As mentioned above, the formation of the off-diagonal algebraic order of O(4) symmetry is impossible at any finite $T$. The situation is changed by the term $\sim (\vec{\psi}^*)^2 (\vec{\psi})^2$ which breaks the O(4) symmetry down to O(2)$\times$Z$_2$. This makes the formation of the algebraic order possible (cf. [\onlinecite{Rubo-07}]).

Introducing real $\vec{a}$ and imaginary $\vec{b}$ parts of $\vec{\psi}=\vec{a} + i \vec{b}$, that is, $\Phi = a_x + ib_x,\, \Phi_\perp = a_y + ib_y$ in Eq.(\ref{Phi}), the uniform term in Eq.(\ref{Upsi}) becomes
\begin{widetext}
\beq
U= -\mu (\vec{a}^2+\vec{b}^2) +
g\Big(3(\vec{a}^2+\vec{b}^2)^2 - 4(\vec{a}\times\vec{b})^2\Big) -\Big((\vec{\eta} + \vec{\eta}^*)\vec{a} + i (\vec{\eta}^*-\vec{\eta})\vec{b}\Big) ,
\label{UMF}
\eeq
\end{widetext}
where $\mu = \mu_0 + \mu_{\perp}$ and, for the sake of generality, the term linear in $\vec{\psi}$ and $ \vec{\psi}^*$ induced by the external pumping $\sim \vec{\eta}$ [\onlinecite{Nature}] has been introduced.

\begin{figure}[!htb]
	\includegraphics[width=1.1 \columnwidth]{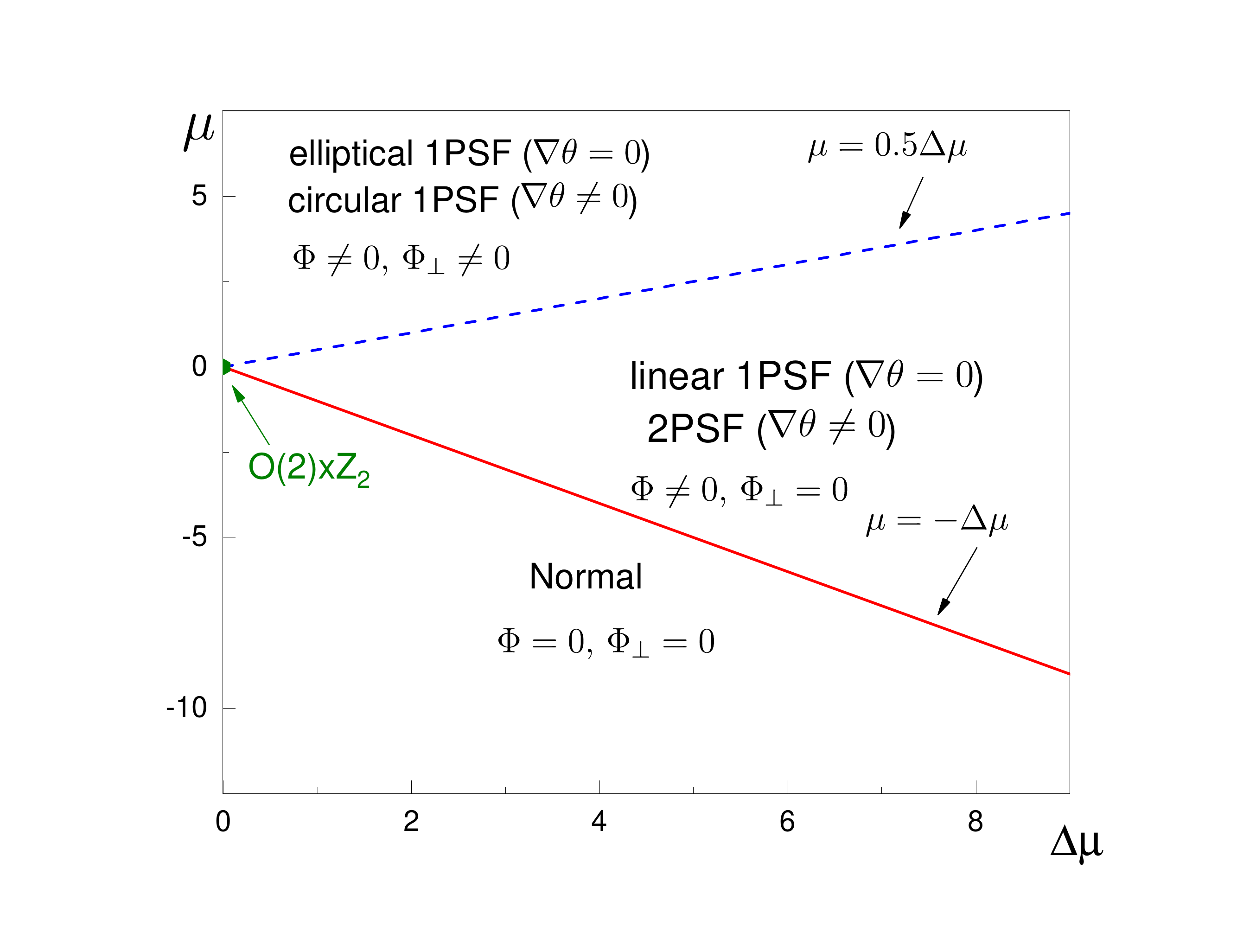}
	\vskip-8mm
	\caption{ Mean field phase diagram ($\vec{\eta}=0$, units are arbitrary). For the notations see the text. }
	\label{fig1}
\end{figure}

At the mean field level the condensation corresponds to $\mu > 0$. The lowest energy of the functional at $\vec{\eta} = 0$  is achieved for $\vec{b} \perp \vec{a}$ and $\vec{a}^2 = \vec{b}^2 = \mu/(8g)$. Explicitly,
 $b_x= a_y$,  $b_y=-a_x$ or $b_x= - a_y$,  $b_y=a_x$ so that
\beq
\psi_x= a_x \pm i a_y,\,\, \psi_y= a_y \mp i a_x,
\label{MF1}
\eeq
where the sign $\pm$ is correlated in both equations and it represents two directions of circularly of the polarized light. Thus, the ground state of the condensed light is characterized by a spontaneous circular polarization -- left or right handed. This corresponds to  $Z_2$ symmetry. Rotation of the vector $\vec{a}$ (together with $\vec{b}$) in the XY-plane implies O(2) (or U(1)) symmetry. Accordingly, the BEC transition belongs to the O(2)$\times$Z$_2$ universality class, which, in general, differs from the Berezinskii-Kosterlitz-Thouless (BKT) scenario characteristic of O(2) symmetry. On the phase diagram, Fig.~\ref{fig1}, this transition corresponds to the point $\mu=\mu_0+\mu_{||}=0, \Delta \mu = \mu_{||}-\mu_\perp=0$ labeled as "O(2)$\times$Z$_2$".

The emerging order is algebraic in 2D because of the gaussian rotational fluctuations of the vectors $\vec{a}$ and  $\vec{b}$ locked to each other. Thus, the one-photon density matrix (OPDM) behaves as $\langle \vec{\psi}^*(\vec{x})\vec{\psi}(0) \rangle \sim 1/|\vec{x}|^{1/K}$ with some index $K > 0$ determined by the parameters and temperature (see in Ref.[\onlinecite{BKT-00}]).

It should also be emphasized that the product $S_z = \vec{a}\times \vec{b} \sim i\vec{\psi}^* \times \vec{\psi}$ represents the order parameter characterized by Z$_2$ symmetry -- similarly to 2D Ising model for the spin variable $S_z \sim \pm 1$. It forms trivially as long as O(2) symmetry is broken.
There is another option -- the O(2)$\times$Z$_2$ transition may proceed as two successive transitions, with the $S_z$ condensing first and O(2) to follow  as temperature lowers further down.

We note that the field theories with U(1)$\times$Z$_2$ symmetry  describe multi-band superconductors exhibiting spontaneous time reversal symmetry breaking [\onlinecite{Babaev_2011,Chubukov,Babaev_2013}]. These feature a variety of topological excitations  [\onlinecite{Babaev_2011}] and also allow for the splitting of the transition into two [\onlinecite{Babaev_2013}]. A significant difference between these systems and the condensate of light lies in the structure of order parameters -- scalars in Refs.[\onlinecite{Babaev_2011,Chubukov,Babaev_2013}] and a complex vector $\vec{\psi}$ for the light. This, in particular, determines the effect which is not present in superconductors -- paired photonic condensate induced by non-Abelian gauge disorder. This effect, which represents the main result of our work, is described below. Here and below we will be using the abbreviation mPSF (m-photon superfluid) to characterize  algebraic order occurring in m-photon (m=1,2) density matrix.

\subsection{Dipolar anisotropy: the uniform case}\label{unif}

Let's consider a situation when there is a residual dipolar anisotropy  characterized by some finite field $\vec{d}_0(\vec{x})$ which is a result of the microscopic averaging of $\vec{d}(x,y,z)$.
It can be induced by external fields  or form spontaneously as a part of the orientational disorder.
Then, the tensor $d_{ij}$ in Eqs.(\ref{dij}) will deviate from the isotropic form (\ref{iso}) so that
 its eigenvalues $\mu_{||},\mu_\perp$ become unequal to each other. Without loss of generality
we will assume that $\Delta \mu=\mu_{||}-\mu_\perp >0$.
Then, the order will form first in $\Phi$  while $\Phi_\perp=0$ in the representation (\ref{Phi}).
As follows from the minimization of $U$ in Eq.(\ref{Upsi}), this occurs in the range $\mu=\mu_0+\mu_\perp < 0$ and $\mu + \Delta \mu>0$.

If $\vec{d}_0 $ is uniform in space, the emerging (algebraic) order is characterized by O(2) symmetry of the 1PSF linearly polarized along this vector (if ignoring $\vec{\eta}$ in Eq.(\ref{UMF})).
This transition is shown in Fig.~\ref{fig1} by the solid line $\mu=-\Delta \mu$. The dashed line, $\mu=0.5 \Delta \mu$,  corresponds to the condensation of the second field, $\Phi_\perp$ proceeding as Z$_2$ transition. As a result, the 1PSF becomes elliptically polarized above this line.

There is also a possibility of a non-uniform one-component condensation if the residual anisotropy $\vec{d}_0$, while persisting on a mesoscale,  self averages to zero on the spatial scale of the cavity. Under this condition the 2PSF phase forms.

\subsection{Non-uniform dipolar anisotropy: Geometrically paired photonic condensate}\label{main}

Now let's consider a situation when $\vec{d}_0$ is not spatially uniform.  We start with the case $|\vec{d}_0 |=$const.
Then, the local eigenvalues $\mu_\|, \mu_\perp$ of $d_{ij}$ are uniform while the orientation of the tensor is set by a non-uniform
$\vec{\nabla}\theta \neq 0$. Since the threshold for the condensation is controlled by the quadratic part of the free energy (\ref{Upsi}), we will ignore anisotropy of the quartic term and will consider it  isotropic as represented in Eq.(\ref{U4P}).

It is important to note that a long wave structure of the field $\vec{\nabla}\theta $ without windings of the angle $\theta$ will not produce any significant deviation from the results discussed above in Sec.~\ref{unif}  as long as the scale of the variations is larger than the healing length.
The most interesting option occurs when  $\vec{\nabla}\theta $ is characterized by topological defects -- vortices. Let's consider a situation when there is a plasma of such proliferated (and frozen in) vortex-antivortex pairs of the field  $\theta(\vec{x})$. If the distance between these vortices $\xi_d$ is larger than the healing length $l_h$, the phase $\Phi \neq 0,\ \Phi_\perp = 0$ (realized in between the two lines in Fig.~\ref{fig1}) persists. Then, setting $\Phi_\perp=0$ in Eqs.(\ref{Upsi}) the free energy becomes
\begin{widetext}
\beq
U= \int d^2x\left[  \frac{\hbar^2}{2m}|\vec{\nabla} \Phi|^2 +  \left(\frac{ \hbar^2(\vec{\nabla}\theta)^2}{2m} -\mu -\Delta\mu\right)  |\Phi|^2 + 3g|\Phi|^4 \right].
\label{HHsm}
\eeq
\end{widetext}
Thus, the gauge effect of the field $\vec{\nabla}\theta$ vanishes. The remaining term $\sim (\vec{\nabla}\theta)^2$ contributes to a weak suppression of the local chemical potential -- as long as $\xi_d >> l_h$.  According to the Harris criterion (see in Ref.[\onlinecite{BK}]), the disorder produced by  $ \sim (\nabla \theta)^2$ in Eq.(\ref{HHsm}) is diagonal and, thus, is irrelevant at the transition marked by the solid line in Fig.~\ref{fig1}. This means that the transition of $\Phi$ remains of the BKT type even in the presence of the disordered field $ \sim (\nabla \theta)^2$.
[The solid line in the phase diagram Fig.~\ref{fig1} shifts slightly upward by the term $\approx \frac{ \hbar^2(\vec{\nabla}\theta)^2}{2m} \sim \hbar^2/(m\xi_d^2)<<\mu$].

It is important to note that, despite the emerging order in the field $\Phi$, the one-photon density matrix (OPDM) becomes  disordered -- as long as $\theta$
contains frozen in windings. These windings are now imprinted onto the physical field
 $\vec{\psi}= \vec{n} \Phi $ through the factor $\vec{n}$. Accordingly, the OPDM becomes
\beq
\langle \psi^*_i(\vec{x}) \psi_j(\vec{x}')\rangle = n_i(\vec{x}) n_j(\vec{x}')  \langle \Phi^*(\vec{x}) \Phi(\vec{x}')\rangle.
\label{one}
\eeq
Since $\vec{n}$ and $\Phi$ are not, practically, coupled, the averaging over the disorder $\langle ... \rangle_{\theta}$  in Eq.(\ref{one}) can be applied to the factor  $ n_i(\vec{x}) n_j(\vec{x}')$ only. This produces exponential behavior $\langle n_i(\vec{x}) n_j(\vec{x}')\rangle_{\theta} \sim \exp(- |\vec{x}- \vec{x}'|/\xi_d)$ and, accordingly, $ \langle \psi^*_i(\vec{x}) \psi_j(\vec{x}')\rangle \sim \exp(- |\vec{x}- \vec{x}'|/\xi_d) \to 0$, as long as $|\vec{x}- \vec{x}'| >> \xi_d$.

In drastic contrast, the two-photon density matrix (TPDM) $ \rho^{(2)}_{ijkl}= \langle \psi^*_i(\vec{x}) \psi^*_j(\vec{x}) \psi_k(\vec{x}') \psi_l(\vec{x}')\rangle$ retains the algebraic order. The simplest way to see this is to consider the scalar product
$\vec{\psi}^2$ which, after taking into account Eq.(\ref{Phi}) becomes $\vec{\psi}^2= \Phi^2$. That is, $ \rho^{(2)}_{iikk}=\langle (\Phi^*(\vec{x}))^2(\Phi(\vec{x}'))^2\rangle \sim |\vec{x}-\vec{x}'|^{-2/K}$.

In general,
\beq
\rho^{(2)}_{ijkl}=\langle n_i(\vec{x}) n_j(\vec{x}) n_k(\vec{x}') n_l(\vec{x}')\rangle_\theta \cdot
  \langle (\Phi^*(\vec{x}))^2 (\Phi(\vec{x}'))^2\rangle .
\label{TPDM}
\eeq
 At distances larger than $\xi_d$ , the averaging over the disorder gives  $ \langle n_i(\vec{x}) n_j(\vec{x}) n_k(\vec{x}') n_l(\vec{x}')\rangle_\theta=\delta_{ij} \delta_{kl}/4$.
At distances $|\vec{x}-\vec{x}'|<\xi_d$ the factor $\langle n_i n_j n_k n_l\rangle_\theta$ becomes $ \sim (\delta_{ij} \delta_{kl} + \delta_{ik}\delta_{jl} + \delta_{il}\delta_{jk})/8$.  Overall, however, $\rho^{(2)}_{ijkl} \propto \langle (\Phi^*(\vec{x}))^2 (\Phi(\vec{x}'))^2\rangle \sim |\vec{x}-\vec{x}'|^{-2/K}$. More specifically,  the component $ \rho^{(2)}_{xxxx}$ changes from
$(3/8)  \langle (\Phi^*(\vec{x}))^2 (\Phi(\vec{x}'))^2\rangle$ at $|\vec{x}-\vec{x}'|<\xi_d$ to $(1/4)  \langle (\Phi^*(\vec{x}))^2 (\Phi(\vec{x}'))^2\rangle$ in the opposite limit.
For the component $\rho^{(2)}_{xxyy}$ the corresponding limits are $(1/8)  \langle (\Phi^*(\vec{x}))^2 (\Phi(\vec{x}'))^2\rangle$ and $(1/4)  \langle (\Phi^*(\vec{x}))^2 (\Phi(\vec{x}'))^2\rangle$. Finally, for the component $\rho^{(2)}_{xyxy}$ the limits are $(1/8)  \langle (\Phi^*(\vec{x}))^2 (\Phi(\vec{x}'))^2\rangle$ and $\to 0$. This behavior describes the 2PSF phase (labeled as " 2PSF $(\vec{\nabla}\theta\neq 0)$" in Fig.~\ref{fig1}).
In this phase the order parameter becomes a complex tensor
$\psi_i \psi_j$, while $\psi_i=0$.

The condensation of the field $\Phi_\perp$ can restore the one-photon order by introducing vortices into the fields $\Phi,\, \Phi_\perp$ which compensate partially the disorder created by $\theta(\vec{x})$.
This mechanism corresponds to the transformation of the phase gradient $\vec{\nabla}\varphi$ of the fields $\Phi$ and $\Phi_\perp$ from being vortex free to acquiring windings as  $\vec{\nabla}\varphi \to  \vec{\nabla}\varphi = \vec{\nabla}\tilde{\varphi} - \vec{\nabla}\theta$, with $\tilde{\varphi}$ denoting the irrotational part of the phase. As a result, the OPDM of the physical field $\vec{\psi}$ acquires algebraic order of circularly polarized light. This phase occurs above the dashed line in Fig.~\ref{fig1} and it is labeled as "circular 1PSF ($\vec{\nabla}\theta \neq 0$)".
More details on the 2PSF to 1PSF transformation are given in the Appendix \ref{AppB}.

Finally, we note that including disorder in the modulus of $\vec{d}_0$  will affect the value of $\Delta \mu$ in Eqs.(\ref{Upsi},\ref{HHsm}).
 However, weak disorder fluctuations of $\Delta \mu$ should not modify the above conclusions -- at least for the O(2) transition (see in Ref.[\onlinecite{BK}]). However, once the disorder becomes strong, a Bose glass phase can emerge.

\section{Phase separation in the TLS-PSF system}\label{sec:PS}

The phase diagram, Fig.~\ref{fig1}, has been constructed within the assumption of frozen in TLS molecules -- translationally and orientationally.
It is possible, however, that at the time of experiment TLS molecules can adjust themselves to the PSF. Below we will show that such an adjustment can lead to the phase separation effect. The reason for this is the negative contribution (\ref{Uwang},\ref{eps}) to the chemical potential $\mu_\xi$ of TLS. Thus, there is a possibility that finite density of PSF will induce attractive collective interaction between TLS molecules which will overcome their entropic tendency toward spreading apart.

Here we will not be discussing the orientational dynamics and will focus on the positional degrees of freedom only.
The density profile of TLS spatial distribution can be described by the entropic contribution to free energy $F_0=Tn_0V\left[\xi\ln \xi + (1-\xi)\ln(1-\xi) - \mu_\xi \xi\right]$, where $n_0$ is a maximum possible density of TLS molecules determined by their typical size ($\sim 1$nm); $\xi$ stands for their fractional density, that is, $\xi=n/n_0<1$; and $V$ is the system volume.

 If PSF is present, the term $U_\psi$, Eqs.(\ref{Uwang},\ref{eps}), must be added to $F_0$ (with the part corresponding to $\vec{\psi}=0$ subtracted).
It is important to notice that $U_\psi \propto n_0 \xi$. Thus, the difference $U_\psi([\vec{\psi}]) -U_\psi([\vec{\psi}=0]) $ can be represented as $  U' n_0 \xi$, where  $U' $ does not depend on $\xi$.
In order to simplify the analysis, we will consider here the case of no residual dipolar anisotropy. Then, $ U'$ can be obtained from Eqs.(\ref{Upsi},\ref{U4P}), where $g$ is defined below Eq. (\ref{iso2}).
Also in Eq.(\ref{Upsi}), $\mu_\| = \mu_\perp$ and both are $\propto n_0\xi$. There is also a contribution $ -\mu_0 n_pV/L_z$ determined by the external pumping of photons, where we set $\Phi = \Phi_\perp =\sqrt{n_p/2}$ as real and uniform, with $n_p$ being the 2D PSF density. Finally, the total free energy (ignoring the gradients) becomes
\beq
F=V\left[Tn_0\left[\xi\ln \xi + (1-\xi)\ln(1-\xi) - \mu_\xi\xi\right]    -\mu_0\frac{n_p}{L_z} + \frac{n_0}{L_z}\xi (-\mu'n_p +g' n_p^2)\right].
\label{F}
\eeq
Here we have introduced the notations $\mu'= \partial \mu_\|/\partial (n_0\xi)>0,\, g'= 2\partial g/\partial (n_0\xi)>0$ for the constants independent of the mean TLS density $n_0\xi$. Using Eqs.(\ref{iso},\ref{iso2},\ref{dij},\ref{dijkl},\ref{c_4}) in the limit $\beta \delta >>1$, one can find
$\mu'\approx d^2\hbar \omega_0/6\varepsilon_0 \delta$ and $g'\approx d^4(\hbar \omega_0)^2/10 \varepsilon_0^2 L_z \delta^3$.

The condition for the phase separation requires presence of two local minima in $F$. The minimization of $F$, Eq.(\ref{F}), with respect to $n_p$ gives
\beq
n_p=\frac{\mu'}{g'}\frac{\xi-\xi_1}{2\xi}\theta_H(\xi-\xi_1),
\label{np}
\eeq
where $ \theta_H(\xi-\xi_1)$ is the Heaviside step function and $\xi_1=- \mu_0/\mu'n_0$. [Since $\xi >0$ and $\mu'>0$, the inequality is nontrivial only if $\mu_0<0$].
Thus, after eliminating $n_p$ the total 3D free energy (dimensionless) density $f=F/(Vn_0T)$ can be represented as
\beq
f= \xi\ln \xi + (1-\xi)\ln(1-\xi) - \mu_\xi \xi -\kappa \theta_H(\xi - \xi_1)\frac{(\xi-\xi_1)^2}{\xi},
\label{f}
\eeq
where  the notation $\kappa=\mu'^2/4g'TL_z \approx 5\delta /72T$ is introduced.

The value of $\xi_1$ determines the condensation transition, so that if $\xi<\xi_1$, the PSF density, $n_p$, vanishes.
At this point it is important to emphasize that $\mu_0 >0$ (that is, $\xi_1<0$) implies instability leading to $f \to -\infty$ in Eq.(\ref{f}), which corresponds to unlimited growth of $n_p$ as $\xi \to 0$ (in the grand canonical ensemble). Thus, we consider this regime with $\mu_0>0$ as unphysical. In contrast, the situation of $\mu_0 $ being strongly negative corresponds to a weak pumping limit. As $\mu_0$ becomes less negative, the density of thermal photons grows and eventually the contribution to the chemical potential $\sim \mu' n_0\xi  $ due to TLSs will lead to zeroing of the total chemical potential, that, is $\xi_1=\xi$ which implies the transition to finite $n_p$.

\begin{figure}[!htb]
	\includegraphics[width=1.1 \columnwidth]{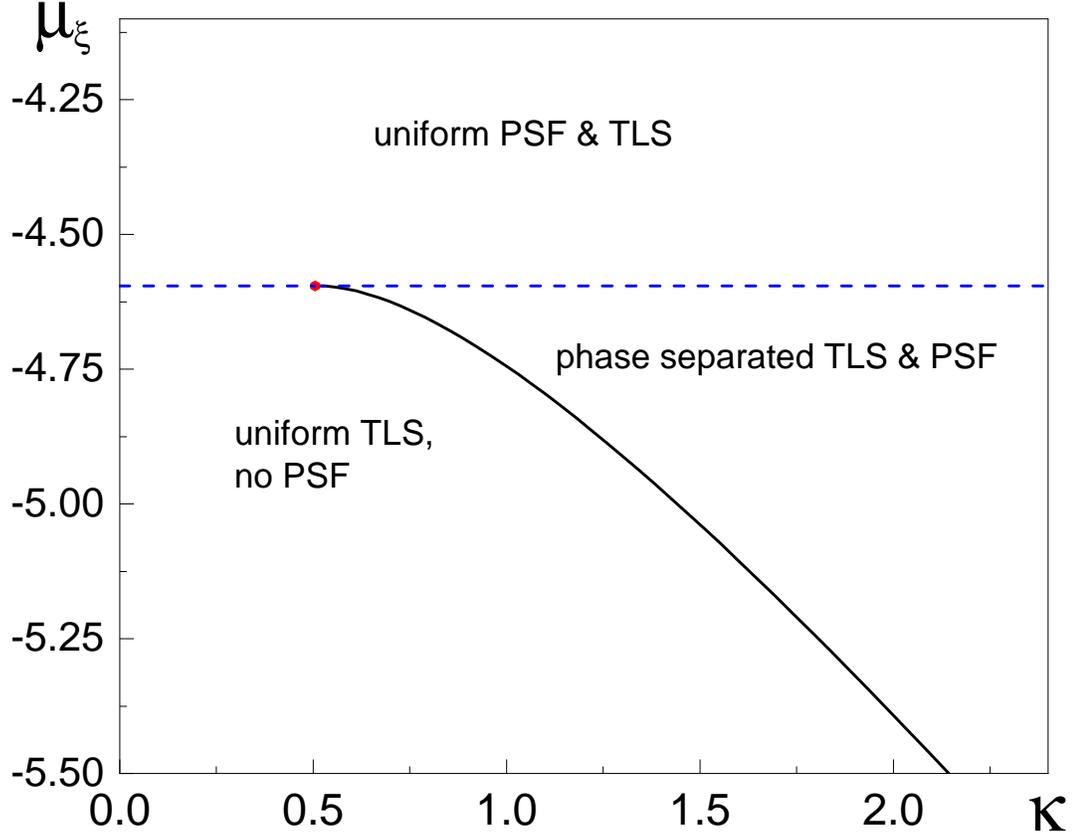}
	\vskip-8mm
	\caption{The phase diagram for the TLS and PSF mixture. The solid and dashed lines mark the Ist and IInd order transitions, respectively. The dot represents the bicritical point. The parameters $\mu_\xi, \kappa$ are defined in terms of the original ones in Sec.\ref{sec:PS}. The value of the parameter $\xi_1=0.01$.  }
	\label{fig2}
\end{figure}

The minimization of $f$ in Eq.(\ref{f}) with respect to $\xi$  leads to the phase diagram shown in Fig.~\ref{fig2}.
It is represented in terms of the parameters $\kappa$ and $\mu_\xi$ for a fixed value of $\xi_1$. This diagram features three phases: i) uniform TLS solution with zero density of the PSF; ii) uniform PSF and TLS solution; iii) phase separated  PSF and TLS. The solid lines mark the position of the corresponding Ist order transitions. The dashed line indicates the IInd order transition. All three phases meet at the bicritical point marked by the solid dot.

\begin{figure}[!htb]
	\includegraphics[width=1.1 \columnwidth]{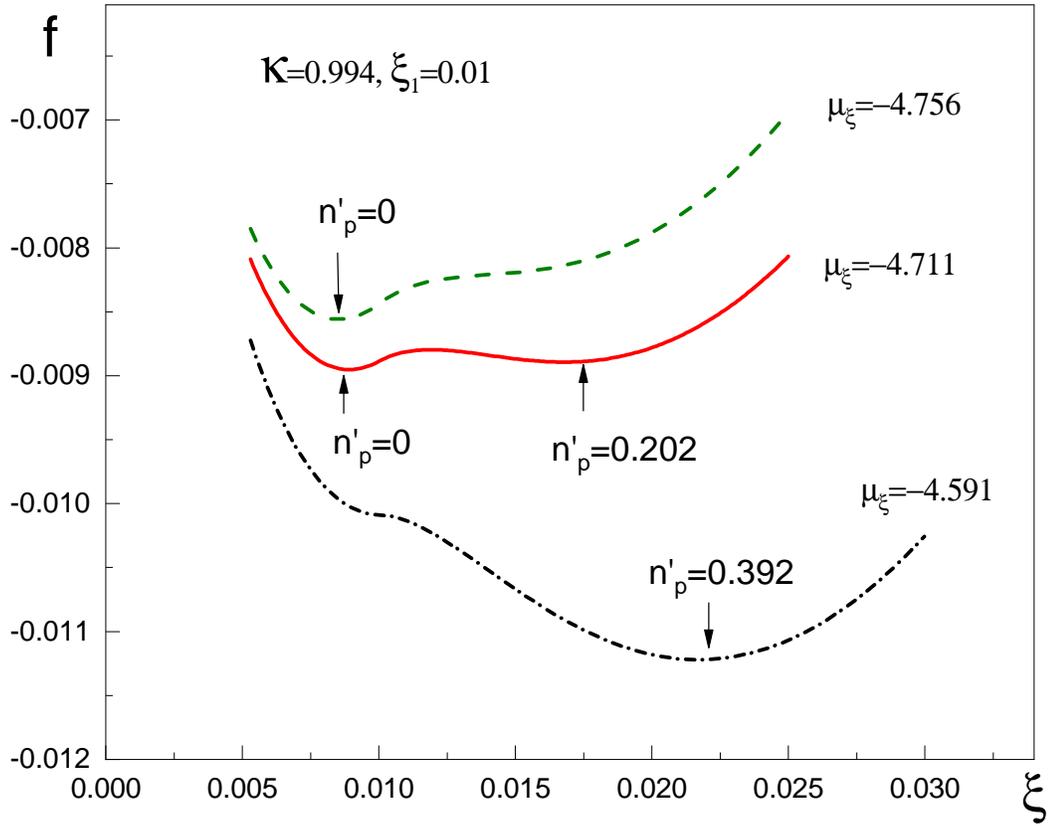}
	\vskip-8mm
	\caption{ Dimensionless free energy density as a function of the fractional TLS density. The upper curve belongs to the uniform TLS phase without any PSF. The middle one is inside the phase separation region and the lower one belongs to the uniform PSF and TLS phase. The corresponding values of $\kappa$, $\mu_\xi$ and $\xi_1$ are shown. The vertical arrows indicate the positions of the minima, with the corresponding values of the dimensionless condensate density $n'_p=n_pg'/\mu'$ shown (as defined in Eq.(\ref{np})).}
	\label{fig3}
\end{figure}

The lower boundary can be found from the  the requirement that the first and the second  derivatives of $f$ with respect to $\xi$ vanish.
Parametrically, it is described by the equations
\beq
\kappa(\xi_c)=\frac{\xi_c^2}{2\xi_1^2(1-\xi_c)}\theta_H(\xi_c-\xi_1),\,\,\,
 \mu_\xi(\xi_c)=\ln\frac{\xi_c}{1-\xi_c}-\frac{\xi^2_c- \xi_1^2}{\xi_c^2} \kappa(\xi_c),
\label{line}
\eeq
where $\xi_c$ (which satisfies $\xi_1 \leq \xi_c <1$) has a meaning of the value of $\xi$ where the inflection of $f(\xi)$ occurs.
The upper phase boundaries are straight  lines parallel to the $\kappa$-axis. They are determined by the condition $\xi=\xi_1, \mu_\xi=\ln(\xi_1/(1-\xi_1))$, with the bicritical point located at $\kappa=1/(2(1-\xi_1)), \mu_\xi=\ln(\xi_1/(1-\xi_1))$.
The illustration of the free energy behavior in all three phases is shown in Fig.~\ref{fig3}.

\section{Discussion}\label{Disc}
Our phenomenological analysis of condensed phases of light points out to multiple possibilities characterized by various symmetries and symmetry breaking patterns of PSF. The algebraic condensation of light in 2D can only occur if the TLS dipolar transition is not fully degenerate.  Then, the resulting symmetry of the condensate is O(2)$\times$Z$_2$, and the algebraic long range order becomes possible. The orientational anisotropy in the TLS ensemble can break this transition into two.

There are two main predictions reported here: i) The geometrically paired photonic superfluid -- that is, 2PSF (according to the suggested nomenclature); ii) Phase separation effect featuring the multicritical point.  The ensuing questions are, first, what are the specifics of these phases for the photonic setups, and, second, how to observe them and the corresponding transitions.

At this point we emphasize that, in contrast to the paired condensates of photons [\onlinecite{Keeling_2PSF}] or ultracold atoms [\onlinecite{Radz}] requiring  an effective attraction between photons, the pairing found here is of purely geometrical nature. It is induced by the non-Abelian disorder in the TLS ensemble, and it does not require any attraction between photons. This effect can be assigned into the class of phenomena named as {\it order by disorder} [\onlinecite{Villain}]. In some sense, there is a hidden algebraic 1PSF order in the non-observable field $\Phi$ while there is no algebraic  order in the observable field $\vec{\psi}$ in the representation (\ref{Phi}).

One option of detecting such a phase would be by controlling
orientation of the dye molecules (by electric or magnetic field) so that $\vec{d}$ of the dipolar transition can be ordered. Then, the non-observable field $\Phi$
 will become observable because $\vec{n}$ in the representation (\ref{Phi}) orders and, thus, restores
 the one-photon coherence in $\vec{\psi} \sim \vec{n}\Phi$.
As discussed above, detecting 2PSF directly requires two-photon correlation spectroscopy.

The phase separation transition discussed here is induced by the 2D algebraic condensation of photons.
It is controlled by several parameters such as fractional concentration of dye molecules $\xi$ and the
photonic chemical potential $\mu_0\sim \xi_1$.
In addition, temperature $T$ and detuning $\delta$ can be varied
for the phase separation to occur. Thus, the system has several "knobs" to adjust.
The phase separation requires a significant time delay $t_{ps}$ to develop -- because TLS centers should diffuse over a typical distance
of the order of the system size $R$ to form a non-uniformity. This time can be estimated as $t_{ps} \sim R^2/D$,
where $D$ is the diffusion coefficient. Using the Stokes-Einstein relation $D=T/(6\pi \eta_v r_0)$, where
$\eta_v \approx $1mPa$\cdot$s is the dynamical viscosity of the solvent (alcohol) and $r_0\sim$ 1nm as a typical size of
dye molecule, the estimate $D\sim 10^2\mu$m$^2/s$ gives the phase separation time scale
as $t_{ps}\sim 1$s for $R\sim 10\mu$m.   Observing such a time delay between starting the external pumping of photons into the cavity
and the formation of the dense photonic cloud will be a "smoking gun" for the effect.

It is worth mentioning that, in the presence of the pumping of light into the cavity (described by the bias $\vec{\eta}$ in Eq.(\ref{UMF})) the phases discussed above can be distorted and then destroyed. A strong external bias $\vec{\eta}$ may essentially lock the condensate polarization along the external field. [This what was likely observed in Ref.[\onlinecite{Greveling}]]. Upon lowering $\eta$ the crossover to the spontaneous polarization should occur. This effect can serve as a measure of the interaction strength between photons.   An estimate of the field strength for a crossover follows from comparing the last terms in Eq.(\ref{UMF}) as $|\vec{\eta}| \approx g n_{ph}^{3/2}$. A different situation occurs in the cases of the dipolar anistropy when the compettion occurs between the residual $\vec{d}_0$ and $\vec{\eta}$. Thus, the phases will be destroyed if $|\vec{\eta}| \geq |\vec{d}_0| \sqrt{n_{ph}}$.

The presented mean field analysis is only a first step toward detailed studies involving realistic geometries. It gives a qualitative assessment of the emerging possibilities. There are several important questions to answer.  One of them is about the nature of the multicritical  point in the phase diagram, Fig.~\ref{fig2}. Its universality is affected by the presence of the BKT transition as well as by the Ising type.  The nature of the 2PSF-1PSF transition involving the compensation of the frozen in vortices in the effective gauge field needs to be explored. To the best of our knowledge, there is no known analogy to this transition.

Another exciting aspect of the system is the possibility of dynamical interaction between PSF and  essentially classical degrees of freedom of dye molecules -- their translational and rotational diffusive dynamics. This poses fundamental questions about how classical diffusive bath affects off diagonal order and dynamics.   Phase separation effect leads to forming regions of higher TLS density which invokes the necessity to consider the near field direct interaction between TLSs as initiated in Ref.[\onlinecite{Sela-14}]. In view of the emerging very promising experimental possibilities [\onlinecite{Weitz2017}] going beyond BEC, these questions acquire much greater significance.

 {\it Acknowledgments}.
We acknowledge useful discussions with Vladimir Yurovsky and Egor Babaev. We also appreciate useful comments by Johnathan Keeling. This work was supported by the National Science Foundation under the grant DMR1720251.

\appendix

\section{Gross-Pitaevskii functional}\label{AppA}

In order to describe (slow) dynamics of the condensed light, the Berry term $S_B$ in the effective action $S=S_B - \int dt U$, where $t$ is real time,  should be constructed. Within the standard GP approach it is
\beq
S^{(0)}_B= \int dt \int d^2x \, \i (\hbar/2) \vec{\psi}^* \dot{\vec{\psi}} + c.c.
\label{So}
\eeq
 This form guarantees the conservation  of photons and leads to the continuity equation $\partial|\vec{\psi}|^2/\partial t +\vec{\nabla} \vec{J}=0$.

Since the actual conservation is given by Eq.(\ref{conserv}),  the Gross-Pitaevskii equation following from $\delta S/\delta \vec{\psi}^*=0$ must feature the continuity equation
\beq
\frac{\partial (|\vec{\psi}|^2 + \rho_{ex})}{\partial t} + \vec{\nabla} \vec{J}=0,
\label{cons}
\eeq
where we have introduced the local density of the excited TLS centers $\rho_{ex}$.

Since TLSs represent fast degrees of freedom, $\rho_{ex}$ can be found within the approximation of local equilibrium with respect to TLS ensemble as
\beq
\rho_{ex}= n \cdot (1+ \langle \sigma_z \rangle )/2,
\label{denp}
\eeq
where the averaging $\langle \sigma_z \rangle$ is done over the ground and excited TLS states in the presence of the condensed field $\vec{E}$ with the equilibrium Boltzmann weight. The diagonalization of $H$ in Eq.(\ref{Hfull}) within this approximation gives
\beq
\rho_{ex}= \frac{1}{2} \left[1 - \frac{\delta \tanh(\beta \epsilon)}{\epsilon}\right] n,
\label{bar}
\eeq
where $\epsilon$ is defined in Eq.(\ref{eps}).

Then, the Berry term consistent with the conservation law (\ref{cons}) follows as $S_B=S_B^{(0)} + S_e$ where
\beq
S_e=\frac{i\hbar}{2} \int dt \int d^2x dz  \frac{(\vec{d} \vec{E}^*)(\vec{d} \dot{\vec{E}})}{|\vec{d} \vec{E}|^2}\rho_{ex} + c.c.
\label{Sb}
\eeq

In order to project the form (\ref{Sb}) into the XY plane, the integration along $Z$-direction must be performed.
This can be done straightforwardly as an expansion in  $1/\delta$.
Expanding $\rho_{ex}$ up to the first term  $\sim |\vec{d}\vec{E}|^2$ and, then, integrating over $Z$,
\beq
\rho_{ex}= \rho_0 + b (\mu_{||} |\Phi|^2 + \mu_\perp |\Phi_\perp|^2)
\label{bar-ex}
\eeq
where $\rho_0= \frac{1}{2} (1- \tanh(\beta \delta)) \int dz n $ and the representation (\ref{Phi}) has been used;
$b=2\delta c_4/c_2$.

Representing
$\Phi=|\Phi|{\rm e}^{i \varphi}$ and $\Phi_\perp=|\Phi_\perp|{\rm e}^{i \varphi_\perp}$ and substituting into Eqs.(\ref{Sb}) and (\ref{bar-ex}) and dropping full time derivative terms, the Gross-Pitaevskii action $S$ can be written as
\begin{widetext}
\beq
S&=&\int dt \int d^2x \Big[ - \hbar \left((1 + b\mu_{||})|\Phi|^2\dot{\varphi} +  (1+b\mu_\perp)|\Phi_\perp|^2 \dot{\varphi}_\perp \right) - U\Big]
\label{Stot}
\eeq
\end{widetext}
where $U$ is given in Eq.(\ref{Upsi}).

\section{Transition 2PSF-1PSF}\label{AppB}
In the case when both fields are condensed, the mean field solution for (\ref{Upsi}) gives
\beq
\Phi = \Phi_1 \exp(\i \varphi),\, \Phi_\perp=\Phi_2\exp(\i \varphi),\,\, \nonumber \\
\Phi^2_1=\frac{2\mu + 3\Delta \mu}{16g},\,\, \Phi_2^2=\frac{2\mu - \Delta \mu}{16g}>0,
\label{MF}
\eeq
which is valid for $\mu > 0.5\Delta \mu$. Here the phase $\varphi$ is to be determined from the minimum of the gradient part
\begin{widetext}
\beq
\Delta U= \int d^2x \frac{\hbar^2}{2m}\left[(\Phi_1 \vec{\nabla} \varphi + \Phi_2 \vec{\nabla}\theta)^2 + (\Phi_2 \vec{\nabla} \varphi + \Phi_1 \vec{\nabla}\theta)^2\right]
\label{grrr}
\eeq
\end{widetext}
of the energy functional (\ref{Upsi}).

Let's assume that the field $ \vec{\nabla} \theta $ has just a single vortex. Then, the minimum of $\Delta U$ can be reached either for $\vec{\nabla}\varphi=0$ (which corresponds to 2PSF) or for $\vec{\nabla}\varphi =- \vec{\nabla} \theta$, which corresponds to circulary polarized 1PSF. Comparison of these two energies leads to the condition
\beq
\frac{\Phi^2_2}{\Phi^2_1} < 7-4\sqrt{3}
\label{cond}
\eeq
for the existence of the 2PSF. Using the solution (\ref{MF}) in Eq.(\ref{cond}) gives
\beq
\mu < \frac{3\sqrt{3} -5}{2\sqrt{3}-3} \Delta \mu \approx 0.423 \Delta \mu,
\eeq
which conflicts with the requirement $\mu > 0.5\Delta \mu$ needed to have $\Phi^2_2 >0$.

Thus, for  $\mu > 0.5\Delta \mu$ the minimum of $\Delta U$ is achieved for $\vec{\nabla}\varphi = - \vec{\nabla} \theta$, that is when the phase of the fields $\Phi, \Phi_\perp$ contains antivortex. In this case the components of the full field $\vec{\psi}$ become
\beq
\psi_x &=& \left[\frac{\Phi_1 + \Phi_2}{2} + \frac{\Phi_1 - \Phi_2}{2}{\rm e}^{2\i \theta}\right] {\rm e}^{\i \tilde{\varphi}}, \\
\psi_y &=& -\i \left[\frac{\Phi_1 + \Phi_2}{2} + \frac{\Phi_2 - \Phi_1}{2}{\rm e}^{2\i \theta}\right] {\rm e}^{\i \tilde{\varphi}},
\label{psixy}
\eeq
where $\tilde{\varphi}$ is a non-winding part of $\varphi$ (accounting for non-topological excitations --  phonons of 1PSF). The terms $\sim \exp(2\i \theta)$ containing the winding field are washed out at large
distances, and the coherent part of the total field $\vec{\psi}$ becomes
\beq
\psi_x = \frac{1}{2}(\Phi_1 + \Phi_2) \exp(\i \tilde{\varphi}), \,\, \psi_y = - \frac{\i}{2}(\Phi_1 + \Phi_2) \exp(\i \tilde{\varphi}),
\eeq
which describes 1PSF with circular polarization.
Such a transformation renders the gauge field $\vec{\nabla}\theta$ irrelevant, and the algebraic part of the OPDM becomes
$\sim  \langle \exp( i [\tilde{\varphi}(\vec{x}) - \tilde{\varphi}(\vec{x}')]) \rangle \sim |\vec{x}-\vec{x}'|^{-1/K}$.

\end{document}